# A review on machine learning for arterial extraction and quantitative assessment on invasive coronary angiograms

Pukar Baral[1], Chen Zhao[2], Michele Esposito[3], Weihua Zhou[4]

1 Department of Applied Computing, Michigan Technological University, Houghton, MI, USA
2 Department of Computer Science, Kennesaw State University, Marietta GA, USA
3 Department of Cardiology, Medical University of South Carolina, Charleston, SC, USA
4 Center for Biocomputing and Digital Health, Institute of Computing and Cyber-systems, and Health Research Institute, Michigan Technological University, Houghton, MI

Corresponding author:
Weihua Zhou, PhD, Tel: +1 906-487-2666
E-mail address: whzhou@mtu.edu
Mailing address: 1400 Townsend Dr, Houghton, MI 49931

## Abstract

*Purpose of Review* Recently, machine learning has developed rapidly in the field of medicine, playing an important role in disease diagnosis. Our aim of this paper is to provide an overview of the advancements in machine learning techniques applied to invasive coronary angiography (ICA) for segmentation of coronary arteries and quantitative evaluation like fractional flow reserve (FFR) and stenosis assessment.

Recent Findings ICA are used extensively along with machine learning techniques for the segmentation of arteries and quantitative evaluation of stenosis, coronary artery disease and measurement of fractional flow reserve, representing a trend towards using computational methods for enhanced diagnostic precision in cardiovascular medicine.

*Summary* Various research studies have been conducted in this field, each using different algorithms and datasets. The performance of these studies largely depends on the algorithms employed and the datasets used for training and evaluation. However, despite the progress made, there remains a need for machine learning (ML) algorithms that can be easily integrated into clinical practice.

# Introduction

Invasive coronary angiography (ICA), also known as X-ray angiograms, serves as the gold standard for assessing the coronary artery anatomy to diagnose coronary artery disease. It provides detailed visualization of coronary arteries through angiographic images or video sequences.

The process of ICA involves catheter insertion into the coronary arteries, followed by contrast injection and X-ray imaging to identify any vessel abnormalities. Then a trained cardiologist examines this ICA for various applications like assessing coronary artery patency, identifying stenosis and localizing lesions and target vessels for coronary revascularization. While other non-invasive procedures like multi-slice spiral computed tomography (MSCT), has been suggested as an alternative to conventional coronary angiography, further studies are necessary to assess its performance in patients with varying levels of coronary artery disease [1].

Accurate segmentation of coronary arteries holds significance beyond its direct application in medical image analysis. It serves as a fundamental step in tasks such as coronary artery disease diagnosis. Additionally, precise segmentation facilitates the extraction of quantitative measures like vessel diameter and stenosis severity and Fractional Flow Reserve (FFR), which are essential for clinical decision-making. In this paper, we aim to explore the research landscape relating to the use of ICA and assess their clinical implications, with a focus on machine learning-based approaches

# Background and related work

We explored scholarly articles for the review paper using keywords like invasive coronary angiography (ICA), deep learning, coronary artery segmentation, X-ray angiograms and machine learning on google scholar. For the scope of this review paper, we are mainly going to focus on those papers that make use of ICA frames or video sequences with machine learning techniques.

### Studies on segmentation of coronary artery tree from ICA

Many studies have been performed, ranging from traditional image processing methods to advanced machine learning algorithms. Traditional image processing methods involve edge detection and region growing to identify the vessel boundaries. With the recent advancement of deep learning, convolution neural networks (CNNs) have gained importance in coronary artery segmentation. Architectures based on CNN can learn complex features directly from the angiography images resulting in a more robust segmentation. The section below summarizes the studies that focus on segmentation of the whole artery tree or a single artery branch from ICA.

### Binary segmentation

Several notable studies have been performed for the task of segmentation of coronary artery tree structure from the ICA frames or videos. Earlier papers often relied on traditional computer vision techniques and manual feature extraction. These papers typically involved preprocessing steps like contrast enhancement, noise reduction and use of enhancement filters followed by segmentation algorithms like thresholding or region growing.

**Region growing**

Region growing algorithms for binary segmentation operate by iteratively expanding from a seed point, incorporating neighboring pixels that satisfy predefined criteria. Initially, a seed point is selected within the region of interest. The algorithm then checks the similarity of neighboring pixels to the seed point based on intensity values or other relevant features. If a neighboring pixel meets the similarity criteria, it is added to the growing region. This process continues iteratively, with newly added pixels becoming seed points for further expansion. However, region growing approaches have limitations, such as reliance on user-supplied seed points and susceptibility to noise and intensity variations, which can lead to incomplete segmentation or over-segmentation.

In [2], authors combined image processing techniques with region growing based segmentation methods for coronary artery segmentation using coronary cine angiograms (CCA), and an accuracy(AC) of 0.9089 was achieved. A median filter and homomorphic filter were applied to improve the vascular structure in the CCA frames. Then an initial mask was created by comparing consecutive frames to find areas where vessels are moving, setting a threshold to remove noisy areas and combining the results. Then this mask was enhanced by removing unwanted areas. Using this mask and CCA images as input in conjunction with several image processing techniques, a new set of images highlighting the vessel regions were produced. Finally, the vessel boundary was detected using contour detection. This technique doesn't rely on any additional prior information regarding vessel regions within the CCA.

In [3] authors presented a multiscale region-growing method (MSRG) for coronary artery segmentation. A region growing rule was developed which integrates the vessel and direction information to the growing rule. Then an iterative multiscale search was performed where the points selected at each stage served as seeds for the later stages. By considering features at different scales and incorporating vessel direction, this method achieved more robust segmentation while dealing with noise, stenosis and low contrast of the angiographic images. The authors evaluated their method on two dataset DS1 and DS2, each consisting of 25 coronary angiography images and achieved an average Dice Similarity Coefficient (DSC) of 0.80 and 0.70 respectively.

In [4] the authors proposed a region growing approach that incorporates a variable vector search area to better adapt to the local vessel shape and capture branches more effectively. Initial seed points were obtained automatically using the Frangi vesselness function. They achieved an average DSC of 0.9133 and average sensitivity (SN) of 0.9171.

**Thresholding-based methods**

Thresholding-based segmentation involves selecting a threshold value to separate pixels into foreground and background based on intensity. Pixels above the threshold are labeled as foreground, while those below are labeled as background, resulting in a binary image. While simple and efficient, selecting an optimal threshold can be challenging, and the approach may struggle with varying illumination or complex backgrounds. Refinement techniques and post-processing steps are often needed to enhance segmentation accuracy.

In [5] authors proposed a two-stage approach to segment coronary arteries from X-ray angiograms. They first used Gabor filters to enhance the vessel structures and suppress the noise. Then, multi-objective optimization was used to find the combination of thresholds based on the defined objectives. Finally, pixels in the images were classified as vessels or backgrounds based on their intensity value relative to the chosen threshold. This method was tested on a test set with 40 angiographic images resulting in accuracy of 0.961.

Similarly, in [6], the authors use filtering and adaptive thresholding to segment the coronary arteries from the angiograms. They evaluated the effectiveness of their proposed method against 13 angiogram images and achieved an average accuracy of 0.9580.

**Graph-based methods**

Graph-based segmentation methods use the graph representation of an image and optimize an energy function to partition the graph into coherent regions, resulting in an accurate segmentation of objects or structures in the image. This section briefly discusses some papers that used graph based approaches for the binary segmentation of arteries from ICA.

Authors of [7] used a super pixel based temporal vessel walker algorithm (SP-TVM) for accurate tracking and segmentation of arteries from angiograms. The first frame was segmented using the seeds given by the user and Vessel Walker algorithm. Then the centerline is calculated, and the vessel of interest is computed using the seeds. A postprocessing step is added to remove any unwanted vessel pixels. Then for remaining frames SP-TVW algorithm is used. At each superpixel size, the algorithm computes the temporal vessel walker using the segmentation result obtained from the previous frame and the result at the first frame. The final segmentation result is computed at a specific superpixel size by combining both SP-TVW results. The results for the same frame at different superpixel sizes are combined to obtain a final segmentation result. The proposed method achieved a mean recall of 0.84 when evaluated against 12 angiographic sequences.

In [8] authors combine multiscale feature analysis with a graph cut approach for the segmentation of coronary arteries. A graph cut algorithm is used to minimize an energy function defined on the graph where the graph is created by assigning each pixel in image as node and connecting neighboring pixels with edges. By incorporating features at different scales this method can potentially capture vessel details at different levels of thickness and improve segmentation accuracy. The authors validated their method on three datasets (DS1, DS2 and DS3) each containing 20, 31 and 40 angiography images respectively and found that it outperforms the other existing methods based on clustering, thresholding and region growing. The proposed algorithm can be computationally expensive in terms of space as the graphs need to be constructed on the fly.

**Multiscale approaches**

Multiscale segmentation involves analyzing images at different resolutions to capture details at varying levels of granularity. By constructing a multiscale representation of the image and performing segmentation independently at each scale, these methods effectively detect objects of different sizes and shapes while minimizing errors caused by noise or artifacts. The segmentation results obtained at different scales are then integrated to produce a comprehensive and accurate final segmentation outcome.

In [9] authors propose a multiscale based method for segmentation of coronary arteries. Three filters are applied to the input image on different scales and for each scale they generate an auxiliary image by getting the maximum values among all filters. These auxiliary images are weighted to get an initial segmentation and the segmentation is further improved by applying an iterative pruning to remove inconsistent elements from the segmentation. The proposed method was evaluated on a test set of 30 angiography images, achieving an average sensitivity of 0.82 in detecting vessel regions.

Authors of [10] combine multiscale analysis with artificial neural networks for the segmentation of arteries in angiography images. The image is analyzed at different scales using Gaussian filters and Gabor filters which are then fed as an input to a multilayer perceptron which gives the detection response. Finally, thresholding method is applied to each pixel to classify each pixel into either vessel or background. The proposed method achieved an area under the Receiver Operating Characteristics (ROC) curve of 0.9775 when evaluated against a test set consisting of 30 angiogram images.

**Deep learning-based methods**

Deep learning for segmentation involves training CNNs to learn features from annotated images, enabling automatic segmentation. The network learns to map input patches to segmentation masks during training, adjusting parameters to minimize the difference between predicted and ground truth masks. Once trained, the CNN segments unseen images by classifying each pixel as foreground or background based on learned features, achieving accurate segmentation of complex structures.

In recent years, there has been a shift towards the adoption of deep learning-based methods for accurately extracting coronary arteries from ICA due to the exponential advancement of deep learning techniques. Unlike traditional approaches that rely on handcrafted features, CNN can automate the task of feature generation allowing for more robust and accurate segmentation of coronary arteries. Also, deep learning-based methods have shown promise in handling the complex and noisy nature of ICA images allowing for extraction of fine vessel structures thus improving the overall quality of segmentation.

In [11] authors used CNNs to segment the arteries from angiography images. The CNN consists of two convolution layers followed by 2x2 and 4x4 max pooling layers respectively followed by a fully connected layer with 2 neurons in the output layer. The contrast of the input image is enhanced and divided into small patches. Then, the CNN is trained on these patches to predict the probability of the center pixel belonging to a vessel or background region. Furthermore, a threshold value is used to classify the pixels belonging to the vessel or background to generate a binary mask. The proposed method achieved an accuracy of 0.935, a sensitivity of 0.90 and a specificity of 0.97 when evaluated against 18 unseen images.

In [12] authors propose a multichannel fully convolution network that takes images before and after the injection of contrast agent to form a multichannel input. Then a U-net like architecture consisting of encode and decoder path is used to generate a final output segmentation map. A hierarchical deformable dense matching method is used to address the displacements between the two images. The proposed method outperformed existing state of the art methods in terms of accuracy, precision, sensitivity, specificity and F1-score when evaluated against ten test images.

In [13] authors propose a deep learning approach based on fully convolutional networks to segment major vessels from the angiography images. Four different deep learning models were evaluated by replacing the backbone of U-Net with a popular network for image classification such as ResNet101, DenseNet121 and InceptionResNet-v2. First, a single frame was selected. The normalization and image augmentation were applied to the ICAs and further fed into the model to generate binary masks. The proposed method achieved a higher F1- score when compared with previous deep learning-based methods with lower segmentation time.

In [14] authors combine CNN with Graph Convolutional Networks (GCNs) into a vessel graph network (VGN) for segmentation of vessels in medical images. Existing methods rely on learning local appearances from images, which doesn't account for overall structure of vessels. To address this problem, the authors of this paper incorporate a GCN into a CNN architecture. This allows the model to jointly use both local features and the global structures to improve the segmentation accuracy. The method when evaluated against a coronary angiography dataset of 3,137 images produced an AUC of 0.9914, suggesting its applicability to the coronary artery segmentation task.

In [15] authors use spatial and temporal information from the angiographic videos for the artery segmentation. The method combines a 3D convolution input layer with a 2D CNN and takes sequences of images as input to produce the segmentation result corresponding to the middle frame. The 3D convolution layer extracts and fuses the temporal information from the 2N+1 frame sequences which is followed by a 2D convolution and max pooling layer. Then a 2D Context Encoder network (CE-Net) which is based on encoder decoder architecture extracts features from the fused feature maps and identifies the foreground and background region of the image. The proposed method achieved better results compared to U-Net and other methods with an accuracy of 0.9855 when N was set as 2.

In [16] employed PSPNet based multiscale CNN to distinguish between coronary artery pixels and background pixels from the angiograms. The method initially extracts features from the input image that is one-eighth the size of the original image. After this the feature maps are fed into the pyramid pooling layer which convolve the feature maps at various sizes and produce the final output. The authors achieved an average accuracy of 0.936 on the test set.

[17] tackles the vessel segmentation problem using the spatio-temporal information using the Spatio Temporal Fully Convolutional Network (ST-FCN). A standard fully CNN captures spatial features within each angiogram frame allowing it to identify vessel-like patterns. The ST-FCNN also includes an 3D convolutional layer, which operates on the complete video sequence. This layer leverages the temporal correlation among successive frames, which enhances the network's grasp of vascular structures in a more holistic manner. The ST-FCN model achieved greater performance compared to the other segmentation techniques, achieving an DSC of 0.90, an AC of 0.92, and a SN of 0.89.

[18] combines deep learning and filter-based features in an ensemble framework to achieve better performance. Filter based features capture vessel like structures using various filters and deep learning features were extracted using common deep learning models selecting the model that achieved best test performance. Then, Gradient boosting decision tree (GBDT) and deep forest classifier were trained on the combined features. The authors reported a mean precision of 0.857, a sensitivity of 0.902, a specificity of 0.992, an F1 score of 0.874, an Area under Receiver Operating Characteristics (AUROC) of 0.947, and an Intersection over Union (IoU) of 0.779 on the GDBT classifier.

[19] proposes a lightweight deep learning network called Bottleneck Residual U-Net (BRU-Net) to address the problem of requiring high performing devices for segmentation. The network architecture is based on U-Net, but incorporates bottleneck residual blocks in the encoder-decoder part to reduce the number of parameters in the network. To account for the reduced complexity and maintain segmentation accuracy, the network incorporates attention modules to model the long-range dependencies in spatial and channel dimensions. They also employed top hat transformation and contrast-limited adaptive histogram equalization (CLAHE) to enhance contrast of arteries in angiograms. The proposed method was trained and tested on two datasets, such as DCA1 and CCA. DCA1 comprises 104 angiograms for training and 30 for testing, while CCA includes 130 angiograms for training and 20 for testing. The proposed method achieved the sensitivity, specificity, accuracy, and an AUC of 0.8770, 0.9789, 0.9729, and 0.9910, respectively for the segmentation task.

[20] proposes a new decoder architecture called EfficientUNet++ which is based on EfficientNet and UNet++ architecture. The network performs segmentation not only for the coronary arteries but also for the catheter present in the angiogram. This provides a reference for scale during analysis, as the catheter dimension is known and also improves the overall segmentation accuracy. The proposed architecture lowers computational demands by substituting UNet++'s blocks with residual bottlenecks using depthwise convolutions. Also, they enhance the bottleneck feature maps using concurrent spatial and channel squeeze and excitation (scSE) blocks. These scSE blocks merge the channel attention of squeeze and excitation (SE) blocks with spatial attention thus improving the performance. They achieved an average DSC of 0.8904 for artery and 0.7526 for catheter segmentation using their best performing models.

[21] proposes an architecture called UNet+++ which is based on U-Net that incorporates several improvements. It makes use of full skip connections which allows the decoder to get detailed information from the early stages of processing, making segmentation more accurate. They incorporate deep supervision so that the model gets feedback not only on the final results but also on the intermediate results to improve its performance. They used a dataset consisting of 616 ICA images to train and test their model to achieve an average dice score of 0.8942.

[22] proposes a novel deep learning architecture based on transformers for the vessel segmentation. Transformers can easily capture long-range dependencies within images unlike CNN. So the proposed method combines CNN with transformers to extract the local and global features from angiography images. The authors incorporate a boundary aggregation module and topology preservation module that uses boundary loss to aggregate boundaries effectively and topological loss to preserve the topology of the vessels respectively. The proposed method was trained and tested on 2,666 angiography images and outperformed other state of the art methods achieving a DSC of 0.9291 and an IoU of 0.8678.

| Reference | Year | Data Type | Dataset | Methodology | Results |
|---|---|---|---|---|---|
| [2] | 2015 | ICA Images | 1,152 ICA Images | Region Growing | Accuracy: 0.9089 |
| [3] | 2016 | ICA Images | 25 ICA images in DS1 and DS2 | Region Growing | DSC: 0.80 and 0.70 in DS1 and DS2 respectively. |
| [4] | 2020 | ICA Images | N/A | Region Growing | DSC: 0.9133 Sensitivity: 0.9171 |
| [5] | 2016 | ICA Images | 40 ICA Images | Thresholding | Accuracy: 0.961 |
| [6] | 2018 | ICA Images | 13 ICA Images | Thresholding | Accuray: 0.9580 |
| [7] | 2016 | ICA sequences | 12 ICA sequences | Graph Based | Mean Recall: 0.84 in the overall dataset. |
| [8] | 2017 | ICA Images | DS1, DS2 and DS3 with 20,31 and 40 ICA respectively. | Graph Cut | N/A |
| [9] | 2018 | ICA Images | 30 ICA images | Multiscale approach | Sensitivity: 0.82 |
| [10] | 2019 | ICA Images | 30 ICA images | Multiscale analysis with neural networks. | ROC: 0.9775 |
| [11] | 2016 | ICA Images | 44 ICA images | Convolution Neural Networks(CNNs) | Accuracy: 0.935 Sensitivity: 0.90 Specificity: 0.97 |
| [12] | 2018 | ICA Images | 130 pairs for training 18 pairs for testing. | FCN | Accuracy: 0.9881 F1-Score: 0.8725 |
| [13] | 2019 | ICA Images | 3,302 ICA images from 2042 patients | Based on Fully Convolutional Networks(FCNs) | F1-Score:0.917±0.103 Re: 0.921±0.112 Pr: 0.918±0.103 |
| [14] | 2019 | ICA and retinal images | 3,137 ICA images | Unified GCN and CNN known as vessel graph network(VGN) architecture. | Average precision of 0.915 on the dataset. |

| Reference | Year | Data Type | Dataset | Methodology | Results |
|---|---|---|---|---|---|
| [15] | 2020 | ICA sequences | 8,835 ICA images and labels | Novel spatial-temporal framework 2D CE-Net(Context Encoder Network for 2D Medical image segmentation) | Accuracy:0.9855 Sensitivity:0.7993 Specificity:0.9939 IOU(vessel):0.7137 IOU(Background):0.9847 |
| [16] | 2021 | ICA Images | N/A | Based on PSPNet architecture. | Accuracy: 0.957 Accuracy: 0.936 Sensitivity: 0.865 Specificity: 0.949 |
| [17] | 2021 | ICA sequences | N/A | Spatio Temporal Fully Convolutional Network(ST-FCN) | DSC: 0.90 Accuracy: 0.92 Sensitivity: 0.89 |
| [18] | 2022 | ICA Images | 130 ICA images | Deep learning and filter-based features ensemble framework | F1- Score: 0.874 AUROC: 0.947 Sensitivity: 0.902 Specificity: 0.992 |
| [19] | 2022 | ICA Images | DCA1: 134 and CCA: 150 ICA images | Lightweight deep learning network called Bottleneck Residual U-Net(BRU-Net) | Sensitivity: 0.8770 Specificity: 0.9789 Accuracy: 0.9729 AUC: 0.9910 |
| [20] | 2022 | ICA Images | 270 ICA images | EfficientUNet++ | Dice: 0.8904 |
| [21] | 2023 | ICA Images | 616 ICA images | U-Net+++ | Dice Score: 0.8942 Sensitivity: 0.8735 |
| [22] | 2024 | ICA Images | 2666 ICA images | Based on UT-BTNet i.e multiscale U-shaped transformer. | Dice: 0.9291 IoU: 0.8687 |

**Studies on segmentation of coronary artery major vessels from ICA**

In this section we focus on those studies that use ICA images or video sequences for the semantic segmentation of arteries from the ICA data. Semantic segmentation involves precisely labeling arteries with the aim to automate and streamline the identification and characterization of coronary artery anatomy. Precise identification and labeling of individual arteries through ICA holds significant importance in evaluating stenosis severity and diagnosing coronary artery disease. [23]

**Skeleton-based approaches**

In [23] proposed a method that combined deep learning with centerline extraction method to build a robust coronary artery segment classifier using a support vector machine. Firstly, a vascular tree was extracted from the angiography image using a deep learning model called Feature Pyramid-UNet++ (FP-U-Net++). FP-UNet++ improves the U-Net++ by incorporating feature pyramids. By applying erosion and dilation operations iteratively, the centerline of the vascular tree is generated. A graph is then created where each artery segment is a link between the two nodes and each node represents an arterial segment. Positional, topological and pixel features are extracted from the segment and a classifier is then trained on these features to perform artery classification. The authors classified the arteries into left main coronary artery (LMA), left anterior descending (LAD) artery, left circumflex (LCX) artery, the first diagonal branch (D1) of the LAD, the first obtuse marginal (OM1) artery, and the ramus intermedius (RI) branch. The proposed method was tested on a dataset of 225 images and achieved a mean AC of 0.7033 for artery classification and a mean IoU of 0.6868 for semantic segmentation.

**Deep learning-based approaches**

In [24] proposed a novel deep learning called T-Net for segmenting the main coronary arteries. It is based on nested encoder-decoder architecture that allows the network to capture features at different scales. The dense skip connections directly link corresponding layers in the encoder and decoder to improve the flow of information in the deeper layers. The optimized version of T-Net achieved a DSC of 0.8897 for the main vessel segmentation.

In [25] authors develop variants of deep learning architecture based on U-Net for major vessel segmentation in X-ray angiography images. The proposed method replaces the backbone of U-Net with ResNet, DenseNet and Residual Attention Network for segmentation. These models were then evaluated on a dataset consisting a total of 3,200 angiography images. Architecture with Residual Attention Network as backbone achieved the highest F1 score of 0.921. However, the proposed method still has difficulty in segmenting thin vessel regions in the ICA images.

In [26] authors propose a novel deep learning architecture called Dual-Branch Multi scale Attention Network (DBMN) to address both segmentation and quantification of the segmented vessels. The DBMN consists of Nested Residual Module that extracts and aggregates multi-level and multi scale features and another attentive regression module that uses a two-phase attention block to capture interactive correlations between separated regions and identify informative regions in the image. The proposed method achieved a

DSC of 0.9160, a precision of 0.9210 and a sensitivity of 0.9130 when evaluated against a dataset of 1,893 images.

In [27] proposed a selective ensemble method that combines multiple deep learning models for segmentation of major arteries from the angiography images. The authors integrated U-Net with DenseNet-121 to use as a base model architecture. Then using five different loss functions, five different models were trained. The predictions from individual models were weighted according to ranks of the models. The weight for each segmentation model was determined based on the quality of prediction masks. An internal dataset of 7,426 angiography images was used to train the model and was validated on 556 external images.

In paper [28] addressed the difficulty in segmenting the arteries at three different granularities by introducing a novel progressive perception learning (PPL) framework. The framework includes three modules to address these difficulties. Context Perception Module (CPM) focuses on the main coronary vessel by capturing semantic dependencies between different coronary segments. Interference Perception Module (IPM) aims to improve feature maps by enhancing foreground vessels and suppressing background vessels. Boundary Perception Module (BPM) uses features at the intersection of foreground and background predictions to highlight vessel boundary details. The proposed framework consists of 5 CPMs, 4 IPMs and 4 BPMs. The proposed method when evaluated against 1085 subjects achieved an overall DSC of 0.9585.

**Graph matching approaches**

In [29], Zhao et al. proposes a method that is based on association graphs and graph matching for segmenting and classifying arteries in angiography images. Unlike other approaches that focus on individual pixels, Association-graph based Graph Matching Network (AGMN) constructs association graphs, representing the relationship between corresponding segments extracted from a pair of angiogram images. Each vertex in the association graph signifies the connection between two arterial segments from the separate images. A graph matching network based on GCN is designed to analyze these association graphs and learn the matching between corresponding artery segments across different angiograms. By learning this correspondence, the network can classify unlabeled segments in one image based on labeled segments on another image. The proposed model achieved an average accuracy of 0.8264, an average precision of 0.8276, an average recall of 0.8264, and an average F1-score of 0.8262 when evaluated against a dataset consisting of 263 ICAs to train and validate the model.

In [30], the authors expand upon their previous work on [29] by introducing a novel edge attention mechanism to efficiently capture the importance of edge connections and feature representations within the graph.

| Reference | Year | Data Type | Dataset | Arteries Labeled | Methodology | Results |
|---|---|---|---|---|---|---|
| [23] | 2021 | ICA Images | 225 ICA images | LMA, LAD, LCX, D1, OM1 and R1 | Centerline Detection | SVM gave best accuracy of 0.7033 |
| [24] | 2020 | ICA Images | 4700 ICA images | LAD, LCX and RCA | FCN | DSC: 0.8897 |
| [25] | 2020 | ICA Images | 3200 ICA images | LAD, LCX and RCA | FCN | Precision: 0.919 Recall: 0.924 F1-Score: 0.921 |
| [26] | 2021 | ICA Images | 1893 ICA images | LAD, LCX, RCA | Based on Dual Branch Multi-Scale Attention Network(DBMAN). | DSC: 0.9160 Precision: 0.9210 Sensitivity: 0.9130 |
| [27] | 2023 | ICA Images | Internal Dataset: 7426 ICA images  External dataset: 556 ICA images | LCA, LAD, LCX | Ensemble of deep learning models | DSC: 0.9307 |
| [28] | 2023 | ICA Images | 1085 ICA images.  985 from in house dataset and 134 and 30 from two public datasets. | RCA, LAD, LCX | Deep learning based Progressive Perception Learning Framework. | Dice: 0.9585 IOU: 0.9255 |
| [29] | 2023 | ICA Images | 263 ICA images | LMA, LCX, LAD, D, OM | Association based graph matching | Accuracy: 0.8264 Precision: 0.8276 Recall: 0.8264 F1-score: 0.8262 |

| Reference | Year | Data Type | Dataset | Arteries Labeled | Methodology | Results |
|---|---|---|---|---|---|---|
| [30] EAGMN | 2023 | ICA Images | 263 ICA images | LMA, LCX, LAD, D, OM | Graph matching with edge attention mechanism | Accuracy: 0.8653 Precision: 0.8656 Recall: 0.8653 F1-score: 0.8643 |

| Reference | Artery Type | Accuracy | Precision | Recall/ Sensitivity | F1-Score | DSC |
|---|---|---|---|---|---|---|
| [23] | LMA | 0.8771 | N/A | N/A | N/A | N/A |
|  | LAD | 0.8398 | N/A | N/A | N/A | N/A |
|  | LCX | 0.7461 | N/A | N/A | N/A | N/A |
|  | D1 | 0.6328 | N/A | N/A | N/A | N/A |
|  | OM1 | 0.5168 | N/A | N/A | N/A | N/A |
|  | RI | 0.8235 | N/A | N/A | N/A | N/A |
| [24] | LAD | N/A | 0.9051 | 0.8824 | N/A | 0.8896 |
|  | LCX | N/A | 0,8484 | 0.8669 | N/A | 0.8514 |
|  | RCA | N/A | 0.9264 | 0.9304 | N/A | 0.9253 |
| [25] | LAD | N/A | 0.921 | 0.918 | 0.919 | N/A |
|  | LCX | N/A | 0.882 | 0.897 | 0.889 | N/A |
|  | RCA | N/A | 0.933 | 0.925 | 0.929 | N/A |
| [26] | LAD | N/A | 0.912 | 0.904 | N/A | 0.907 |
|  | LCX | N/A | 0.907 | 0.892 | N/A | 0.898 |
|  | RCA | N/A | 0.943 | 0.941 | N/A | 0.940 |
| [27] | N/A | N/A | N/A | N/A | N/A | N/A |
| [28] | LAD | N/A | N/A | N/A | N/A | 0.9576 |
|  | LCX | N/A | N/A | N/A | N/A | 0.9445 |
|  | RCA | N/A | N/A | N/A | N/A | 0.9746 |
| [29] | LMA | 0.9956 | 0.9911 | 0.9956 | 0.9933 | N/A |
|  | LCX | 0.8432 | 0.8476 | 0.8432 | 0.8452 | N/A |
|  | LAD | 0.8046 | 0.8256 | 0.8046 | 0.8143 | N/A |
|  | D | 0.7956 | 0.7536 | 0.7956 | 0.7736 | N/A |
|  | OM | 0.7565 | 0.7613 | 0.7565 | 0.7539 | N/A |

| Reference | Artery Type | Accuracy | Precision | Recall/ Sensitivity | F1-Score | DSC |
|---|---|---|---|---|---|---|
| [30] | LMA | 1.0000 | 0.9841 | 1.0000 | 0.9919 | N/A |
|  | LAD | 0.8905 | 0.8692 | 0.8905 | 0.8795 | N/A |
|  | LCX | 0.8831 | 0.8444 | 0.8831 | 0.8629 | N/A |
|  | D | 0.8142 | 0.8335 | 0.8142 | 0.8233 | N/A |
|  | OM | 0.7597 | 0.8472 | 0.7597 | 0.7999 | N/A |

**Studies on stenosis evaluation from X-ray angiograms**

In this section, we focus on those studies that utilize coronary angiography images or video sequences for stenosis assessment. Stenosis evaluation usually involves assessing the degree of narrowing in coronary arteries and is also important for diagnosing coronary artery disease, and also to determine appropriate treatment strategies.

**Radius based stenosis evaluation**

Radius-based stenosis detection from ICA involves identifying regions within blood vessels where the diameter is reduced, indicating a potential blockage or narrowing. Stenosis is detected when the ratio of the diameter at the stenotic region to the reference diameter falls below a predefined threshold, indicating a narrowing of the vessel lumen. This method enables automated detection of stenoses from ICA images, facilitating early diagnosis and intervention in cardiovascular disease.

In [31] authors proposed an innovative framework to first extract coronary arteries using deep learning and detect stenosis from the extracted arterial tree using an algorithm that extracts centerlines. The proposed method combines the power of U-Net++ and feature pyramid to develop a FP-U-Net++ network for coronary artery segmentation. Following this, the centerline of the artery tree is extracted, and a diameter is calculated for every pixel in the centerline. Then an edge linking algorithm is applied to detect the junction point and end points to separate the vascular tree into different segments. Then the local minimal is extracted based on the second derivative, where the value of 0 represents the stenotic RoI of the artery. The calculation analyzes each segment to detect stenosis, and separates the stenosis levels into minimal, mild, moderate or severe levels.

**Object detection-based methods**

Object detection-based methods for stenosis detection involve training ML models to automatically identify stenotic regions within ICA images. These methods typically employ annotated datasets to train models to recognize stenosis based on characteristic features such as shape, texture, and intensity patterns. During training, the model learns to differentiate between stenotic and non-stenotic regions by analyzing image patches or entire images. Once trained, the model can efficiently detect stenoses in new images by predicting bounding boxes or segmentation masks around the detected regions. With the advancements in deep learning techniques, these methods can achieve high levels of accuracy and robustness, enhancing their utility in clinical practice for diagnosing cardiovascular disease.

Authors of [32] suggest an architecture named DeepCADD that is based on Mask R-CNN architecture. It uses a ResNet-50 backbone initialized with ImageNet and is trained on synthetic artery segments dataset to improve the feature extraction process. The classifier module contains a classification, bounding box and segmentation in parallel. This module is trained on 118 angiography images and tested on 14 angiography images.

In [33] authors proposed a transformer-based method to make use of the spatio-temporal information presented in the angiography sequences for stenosis detection. Firstly, a feature and proposal extraction module spatial feature maps and proposals from each frame in angiography sequence. It then employs proposal shifted spatio-temporal tokenization (PSSTT) to gather spatio-temporal region of interest features. PSSTT tokenizes the angiography sequences which allows the extraction of visual tokens within a local window. A transformer-based feature aggregation (TFA) network then takes the extracted tokens and uses transformer to learn long-range spatio-temporal context. Finally, a multitask outputs (MTO) module predicts the stenosis bounding box and its classification from the features extracted.

In [34] authors propose a novel object detection network known as Stenosis-DetNet. Firstly, feature maps are extracted from the angiography image to generate the potential candidate boxes. Then, the candidate boxes and feature maps go through classification and regression module to acquire stenosis detection result. The proposed method was trained and tested against a dataset of 166 X-ray sequences and achieved a precision and sensitivity of 0.9487 and 0.8222, respectively.

**Deep learning-based approaches**

Deep learning-based approaches for stenosis detection leverage neural networks to automatically learn features from ICAs, such as angiograms or coronary computed tomography angiography (CCTA) scans. These models typically consist of multiple layers of interconnected nodes that process the input data hierarchically, extracting increasingly complex features. Initially, the model is trained on a dataset of annotated medical images, where each image is labeled with information about the presence and severity of stenosis. Once trained, the model can analyze new images and identify regions suspected of stenosis based on learned patterns and features.

In [35] authors first develop an algorithm that automatically extracts key frames from the angiography sequences to reconstruct the original dataset. Then a deep neural network based on Inception-V3 is trained on the reconstructed dataset to classify stenosis. Self-attention modules were also used to focus on important features in the images. Finally, the obtained results were visualized using heatmap which indicated an important region in the images responsible for classification.

In [36] authors proposed a fully automated approach to estimate stenosis severity form coronary angiograms. The dataset consisted of 1,418,297 image frames and the 70-30 split was used for training and testing for all algorithms except algorithm 3. They developed four neural networks to perform sequential tasks for localization and estimation of the stenosis. Algorithm 1, which was based on Xception, was responsible for the identification of angiographic projection angle. The CNN was initialized with ImageNet weights and all layers were trained using the available dataset. Then algorithm 2, which also used Xception architectures, was initialized with weights trained from Algorithm 1 to identify the anatomical structures in the frames. Frames that contained only Left or Right coronary artery were then passed to Algorithm 3 for further analysis. Algorithm 3, which was based on RetinaNet architecture, was trained to localize the objects in the angiograms. The output of algorithm 3 was bounding box coordinates for any objects present in the angiogram. The algorithm 3 not only localized stenosis but also other objects like guidewires. Finally, algorithm 4 was developed to estimate the severity of stenosis from the input image that is cropped at the previously identified stenosis artery segments. Algorithm 4 was based on modified Xception architecture to predict stenosis severity as a continuous value. The weights of this model were initialized using the weights from Algorithm 1. The authors also employed explainability methods to assess the prediction made by their algorithm.

In [37] authors expand upon their previous work on [36] by developing a pipeline to identify and estimate stenosis severity percentage using angiography videos. The model was based on Swin3D model and was trained and validated on a dataset consisting of 182,418 angiographies. Six different algorithms were developed where data flows sequentially from one algorithm to another. Algorithm 1 and 2 were similar to [36] and the rest of the algorithms from 3-6 were novel to this paper. Algorithm 3 aligns the stenosis bounding box detected by algorithm 2 to 17.5mm by 17.5 mm using spatial translations. This outputs a video containing a stenosis box in all frames as identified by algorithm 2. Algorithm 4 then categorizes the coronary artery into proximal, middle, and distal segments. Algorithm 5 evaluates the content of the resized stenosis box in each frame, focusing on pixels within the central region of the reference area, matched to the predicted coronary artery segment by Algorithm 4. Lastly, a stenosis severity prediction algorithm (Algorithm 6), utilizing a modified Swin3D 21 transformer model, quantifies stenosis severity from the aligned video to predict a continuous percentage in targeted artery segments.

**Studies on fractional reserve flow estimation from ICA**

FFR has proven to be a valuable tool in assessing coronary artery stenosis, aiding in the identification of obstructions that impede oxygen delivery to the heart muscle or induce myocardial ischemia [38]. FFR measures the ratio of the maximum blood flow in a stenotic (narrowed) artery to the maximum blood flow if the artery were normal. However, wire-derived FFR (FFRwire) involves invasive procedures that necessitate the insertion of a separate catheter, the administration of systemic blood thinners, and the placement of a 0.014-inch pressure wire into the affected coronary artery. FFRwire is performed at the time of ICA [39].

**FFR estimation using computational fluid dynamics**

Computational Fluid Dynamics (CFD) based methods for FFR estimation simulate blood flow dynamics within the coronary arteries using mathematical equations and computational algorithms.

In [40] proposed a method to evaluate the performance of ICA derived FFR, i.e $FFR_{angio}$, in patients with complex coronary artery disease involving multiple blockages. The $FFR_{angio}$ involved offline analysis of ICA DICOM images uploaded to a secure server and conducted by an independent operator, who was blinded to the results of the wire based FFR evaluations. Based on these results, the Angio-based fSS (functional severity score) for each patient was computed. Primary endpoints of the study focused on the diagnostic accuracy of $FFR_{angio}$ per lesion and per patient compared to wire based FFR, with sensitivity and specificity determined using a dichotomously scored $FFR_{angio}$ index. Secondary endpoints included correlation analyses between continuously scored FFR values ($FFR_{angio}$ and $FFR_{wire}$) and assessment of the time required for $FFR_{angio}$ measurement compared to wire based FFR. Statistical methods included Pearson's correlation coefficient, Bland–Altman analyses for agreement assessment, and calculation of sensitivity, specificity, and accuracy of $FFR_{angio}$ compared to the wire based FFR. Results from fifty enrolled patients with a total of 118 lesions revealed promising diagnostic accuracy for $FFR_{angio}$, with a mean $FFR_{angio}$ value of 0.81. The study demonstrated that $FFR_{angio}$ has the potential to serve as a reliable alternative to invasive physiological assessment methods for evaluating coronary artery disease severity.

In [41] authors investigated a new non-invasive method for assessing the diagnostic performance of three dimensional quantitative coronary angiography based vessel FFR known as vFFR compared to wire based FFR i.e FFRwire. The vFFR method involved offline analysis of angiographic images and hemodynamic data by a blinded core laboratory, including computation of vFFR using three two-dimensional images to reconstruct coronary arteries and locate the FFR pressure wire. Derived parameters were used to compute vFFR, with vessel geometry derived from validated 3D reconstructions. For endpoints, vFFR recordings were assessed offline for diagnostic accuracy in identifying physiologically significant coronary stenoses, with sample size calculations ensuring statistical power. Statistical analysis included correlation assessments, Bland-Altman plots, ROC curves, and sensitivity/specificity calculations. Enrollment from multiple global sites resulted in 334 patients for final analysis. Results demonstrated promising diagnostic accuracy of vFFR in identifying physiologically significant coronary stenoses, with a sensitivity of 0.90 and specificity of 0.77 for detecting FFR ≤0.80, consistent with previous findings.

In [42] the authors proposed a method to estimate the FFR using ICA. The proposed method estimated the FFR using anatomical features that were extracted from ICA and two physiological features i.e heart rate and blood pressure. Firstly, two angiographies showing the lesion from different angles are selected and are processed to generate a 3D model of the coronary artery using commercially available software. After this, the centerline of the vessels in the selected frames is traced manually and vessel contours are detected and corrected if necessary. A 3D model of the coronary vessel segment containing the lesion is then automatically generated using the delineated vessel contours and other geometric information about the lesion. Personalized boundary conditions, including physiological features and vessel cross-section of the non-diseased proximal segment, along with Murray's law for flow distribution and estimates of coronary microvascular resistance, are inputted into CFD modeling software. This software then simulates changes in arterial pressure and coronary flow under normal and maximal hyperemic conditions. Using these simulations, FFR values are calculated for the entire modeled coronary vessel segment based on the computed pressures at specific locations of interest and at the inlet. The obtained value using CFD modeling was then compared with $FFR_{wire}$. Out of the 100 coronary lesions, 29 were hemodynamically significant i.e with $FFR_{invasive} \leq 0.80$, the calculated FFR values correctly identified these lesions with 0.90 accuracy, showing a sensitivity of 0.79 and a specificity of 0.94.

In [43] authors proposed a method based on rapid flow analysis for the estimation of FFR using ICA images. The 3D structure of the coronary artery tree is constructed from the 2D ICA images using epipolar ray tracing along with other constraints that enforces the tree structure. A compensation mechanism is introduced where the 3D engine uses all available projections to compensate for the displacements due to breathing and patient movements. Then, each branch in the entire 3D tree is analyzed scanning for narrowed regions diameter. A lumped model is then built considering the contribution of each narrowing to the total resistance for flow. Using inlet and outlet boundary conditions, the solution of the lumped model solution gives the estimate of flow rate ratios between stenosed and healthy coronary arteries.

In [44] authors propose a method for estimating FFR using ICAs. The $FFR_{angio}$ method involves several steps. First, at least three DICOM angiograms are acquired and transferred directly to the $FFR_{angio}$ console. Then, the $FFR_{angio}$ system creates a 3D reconstruction of the coronary tree based on the angiograms, using proprietary software and enforcing the tree's structure with epipolar ray tracing and mathematical constraints. Next, the coronary arterial network is modeled as an electric circuit, with each segment acting as a resistor, and the resistance of each vessel is estimated based on its length and diameter. The contribution of each vessel to flow is determined based on its impact on overall resistance, and normal maximal flow is estimated based on the volume of coronary vessels and total coronary length. $FFR_{angio}$ is then calculated as the ratio of the maximal flow rate in the stenosed artery compared to the maximal flow rate in the absence of the stenosis, using the modeled coronary arterial network. Finally, the $FFR_{angio}$ report is sent to the $FFR_{angio}$ core laboratory for review, with the core laboratory blinded to the FFR measurement obtained from the pressure wire sensor. The authors found that $FFR_{angio}$ achieved a high sensitivity, specificity and accuracy when compared with $FFR_{wire}$.

**Machine learning-based methods**

In [45] authors developed a supervised machine learning algorithm to classify lesions as having FFR ≤ 0.80 or > 0.8. They utilized angiographic lumen diameter measurements and clinical features using XGBoost for the classification. The model achieved a sensitivity of 84%, a specificity of 0.80 and an overall accuracy of 0.82 in the test data when 12 high ranking features were used.

In [46] authors proposed an approach to estimate patient specific coronary flow semi automatically using the change in centerline information in different frames of ICA. Firstly, the vascular tree is traced from the angiograms using image preprocessing and Gabor filters. Then, centerline was extracted for the vessel of interest for each frame of the angiogram. Finally, the change in the length of the centerline is used to compute flow velocity, providing patient specific quantitative flow ratio.

**Studies on the measurement of left ventricular ejection fraction**

Left ventricular ejection fraction (LVEF) is a measure used to assess the pumping ability of the heart's left ventricle, which is responsible for pumping the oxygen-rich blood to the body. A reduced LVEF of below indicates impaired heart function, which can be associated with conditions such as heart failure, coronary artery disease, or cardiomyopathy.

In [47] authors proposed a method that combined 3D CNNs and transformers to estimate LVEF using ICA of two left coronary arteries. The proposed method used 3D CNN which was based on ResNet152 architecture and a TimeSformer under the transformer architecture, which was originally developed for video classification tasks. Each of the two images were then fed into both models and their outputs were combined to create a new feature vector. This newly created feature vector was then inputted into a MLP with a single hidden layer to predict the output. The MLP classified the LVEF as positive i.e LVEF ≤ 40% or negative i.e LVEF >40%, which was evaluated against Echocardiogram as the gold standard LVEF. The proposed method was evaluated against a collection of 18,809 angiograms which was divided into training, validation and test set in 70-10-20 split. It achieved an AUC of 0.87, a sensitivity of 0.77 and a specificity of 0.80 in the test set.

In [48] the authors proposed an approach to predict LVEF from ICAs of left coronary artery using a deep neural network called CathEF. CathEF was based on X3D architecture that accepted ICA videos and provided an output as a continuous LVEF percentage value. For all LCA videos from the same study, the DNN's prediction was averaged to get the final study-level performance. The author also evaluated the DNN's by averaging predictions from videos captured at three specific projection angles which are common echocardiographic views used in determining LVEF. The researchers also applied GradCAM, a technique that highlights the regions of the video that contributed most to the DNN' s performance. The performance of the method was evaluated against test data consisting of 813 studies and external validation data consisting of 776 studies. The model achieved an AUROC of 0.911 when considering all left coronary artery (LCA) videos per study. When focusing on LCA videos from three specific projection angles, the AUROC for the test data was 0.908. Similarly, for external validation data, the model achieved an AUROC of 0.906 for all available LCA videos per study and an AUROC of 0.905 when considering LCA videos from the three projection angles.

| Reference | Year | Data Type | Dataset | Methodology/Overview | Results |
|---|---|---|---|---|---|
| [31] | 2021 | Images | 314 ICA images. | Uses FP-U-Net++ to extract arteries from ICA followed by an algorithm that extracts centerlines, calculates the diameters and measures the stenosis levels. | TPR: 0.6840 PPV: 0.6998 |
| [32] | 2022 | Images | 2000 artery segment dataset. 132 Images | Based on masked R-CNN based architecture named DeepCADD. | Sensitivity:0.8913 Precision: 0.8300 |
| [33] | 2023 | ICA Sequences | 233 ICA sequences. | Transformers and CNN | F1 Score: 0.9088 |
| [34] | 2021 | ICA Sequences | 166 ICA sequences | CNN | Precision:0.9487 Sensitivity:0.8222 |
| [35] | 2021 | ICA clips | 452 ICA frames | CNN | Clipwise accuracy of: 0.971 Framewise accuracy of: 0.934 Average clipwise accuracy of 0.965 |

| Reference | Year | Data Type | Dataset | Methodology/Overview | Results |
|---|---|---|---|---|---|
| [36] | 2023 | ICA Images | 84153 Images | Four different algorithms based on CNN | AUC: 0.862 and AUC: 0.869 in external validation dataset. |
| [37] | 2023 | ICA Sequences | 182,418 ICA sequences | Six different algorithms based on CNN. | AUROC: 0.8294 |
| **FFR** | | | | | |
| [44] | 2019 | ICA Images at different projection angles. | 301 subjects | 3D reconstruction of artery tree | Accuracy: 0.9200 |
| [40] | 2019 | ICA Images at different projection angles. | 50 subjects with 118 lesions | Rapid flow analysis of a dynamically derived lumped model. | Accuracy: 0.8400 |
| [41] | 2022 | ICA Images at different projection angles. | 334 patients | 3D- CFD simulation | AUC: 0.9300 |
| [45] | 2019 | ICA Images | 1204 and 297 ICA Images for training for testing respectively | XGBoost | Accuracy:0.8500 |
| [46] | 2019 | ICA Sequences | 328 vessels with FFR data | Change in centerline method | Accuracy: 0.9320 |
| [42] | 2016 | ICA Images at two view angles | N/A | Computational Fluid Dynamics | Accuracy: 0.9000 Sensitivity: 0.7900 Specificity: 0.9400 |
| [43] | 2017 | ICA Images at multiple view angles | N/A | Lumped Model solution | Accuracy: 0.9300 Sensitivity: 0.8800 Specificity: 0.9500 |
| **LVEF** | | | | | |
| [47] | 2023 | ICA Images at two view angles | 18,809 angiograms | 3D CNN + Transformers | AUC:0.87, Sensitivity:0.77 Specificity:0.80 |

| Reference | Year | Data Type | Dataset | Methodology/Overview | Results |
| --- | --- | --- | --- | --- | --- |
| [48] | 2023 | ICA Sequences | 3960 LCA angiography videos from different projection angles | DNN based on X3D architecture | AUROC: 0.908 |

**Conclusion**

The research that we explored in this paper dives deep into the use of coronary angiography images and videos in conjunction with machine learning and deep learning. The integration of machine learning with ICAs has shown promising potential in advancing cardiovascular diagnostics and treatment planning. There is a need to focus on the development of more robust and generalizable algorithms that can handle variations in angiograms. Efforts should be directed towards improving the interpretability and explainability of ML models that can aid clinicians in understanding the reasoning behind model predictions, thereby fostering trust and adoption. However, the progress in this field is significantly held back by the scarcity of high-quality, publicly available datasets. This hinders the development and benchmarking of robust algorithms that could otherwise lead to substantial improvements in patient outcomes. Recent efforts have introduced datasets specifically designed for the segmentation of the coronary arteries and stenosis detection, offering a valuable resource for further advancing research in this area. Therefore, future research efforts should focus on conducting prospective clinical studies and addressing regulatory challenges to facilitate the translation of these technologies into routine clinical practice.